\input colordvi
\input epsf
\catcode`@=11                                   
\catcode`\|=12                                  
\catcode`\&=4                                   

\newcount\ncols         \ncols=\z@              
\newcount\nrows         \nrows=\z@              
\newcount\curcol        \curcol=\z@             
     
\newdimen\thinsize      \thinsize=0.6pt         
\newdimen\thicksize     \thicksize=1.5pt        

\newif\iftableinfo      \tableinfotrue          
\newif\ifcentertables   \centertablestrue       
%
%
     
\let\plaincr=\cr                        
\let\plainspan=\span                    
\let\plaintab=&                         
\let\lparen=(                           
\let\NX=\noexpand                       

     
\def\ruledtable{\relax                          
    \@BeginRuledTable                           
    \@RuledTable}


\def\@BeginRuledTable{
   \ncols=0\nrows=0                             
   \begingroup                                  
    \offinterlineskip                           
    \def~{\phantom{0}}
    \def\span{\plainspan\omit\relax\colcount\plainspan}
    \let\cr=\crrule                             
    \let\CR=\crthick                            
    \let\nr=\crnorule                           
    \let\|=\Vb                                  
%
%
    \ifx\tablestrut\undefined\relax             
    \else\let\tstrut=\tablestrut\fi             
    \catcode`\|=13 \catcode`\&=13\relax         
    \TableActive                                
    \curcol=1                                   
%
%
    \ifdim\tablewidth>-\maxdimen\relax          %
      \edef\@Halign{\NX\halign to \NX\tablewidth\NX\bgroup\TablePreamble}%
      \tabskip=0pt plus 1fil                    
    \else                                       %
      \edef\@Halign{\NX\halign\NX\bgroup\TablePreamble}%
      \tabskip=0pt                              
    \fi                                         %
%
%
    \ifcentertables                             
       \ifhmode\vskip 0pt\fi                    
       \line\bgroup\hss                         
    \else\hbox\bgroup                           
    \fi}


\long\def\@RuledTable#1\endruledtable{
   \vrule width\thicksize                       
     \vbox{\@Halign                             
       \thickrule                               
       #1\relax                                 
       \tstrut                                  
       \plaincr\thickrule                       
     \egroup}
   \vrule width\thicksize                       
   \ifcentertables\hss\fi\egroup                
  \endgroup                                     
  \global\tablewidth=-\maxdimen                 
  \iftableinfo                                  
      \immediate\write16{[Nrows=\the\nrows, Ncols=\the\ncols]}%
   \fi}
     

\def\TablePreamble{
   \linecount                           
   \TableItem{####}
   \plaintab\plaintab                   
   \TableItem{####}
   \plaincr}


\def\@TableItem#1{
   \hfil\tablespace                             
   #1\relax                                     
   \tablespace\hfil                             
    }%

\def\@tableright#1{
   \hfil\tablespace\relax               
   #1\relax                             
   \tablespace\relax}

\def\@tableleft#1{
   \tablespace\relax                    
   #1\relax                             
   \tablespace\hfil}

\let\TableItem=\@TableItem              
     
\def\RightJustifyTables{\let\TableItem=\@tableright}
\def\LeftJustifyTables{\let\TableItem=\@tableleft}
\def\NoJustifyTables{\let\TableItem=\@TableItem}

\def\LooseTables{\let\tablespace=\quad}
\def\TightTables{\let\tablespace=\space}
\LooseTables                                    

%

\newdimen\tablewidth    \tablewidth=-\maxdimen  


\def\setRuledStrut{
   \dimen@=\baselineskip                        
   \advance\dimen@ by-\normalbaselineskip       
   \ifdim\dimen@<.5ex \dimen@=.5ex\fi           
   \setbox0=\hbox{\lparen}
   \dimen1=\dimen@ \advance\dimen1 by \ht0      
   \dimen2=\dimen@ \advance\dimen2 by \dp0      
   \def\tstrut{\vrule height\dimen1 depth\dimen2 width\z@}%
   }%

\def\tstrut{\vrule height 3.1ex depth 1.2ex width 0pt}


\def\bigitem#1{
   \setbox0=\hbox{#1}
   \dimen1 =\ht0 \dimen2 =\dp0                  
   \dimen@ =\baselines@ve                       
   \advance\dimen@ by-\normalbaselineskip       
   \ifdim\dimen@<.25ex \dimen@=.25ex\fi         
   \advance\dimen1 by \dimen@                   
   \advance\dimen2 by \dimen@                   
   \vrule height\dimen1 depth\dimen2 width\z@   
   \copy0}

     
%

     
\def\nextcolumn#1{
   \plaintab\omit#1\relax\colcount              
   \plaintab}
     
\def\tab{
   \nextcolumn{\relax}}


\def\vb{
   \nextcolumn{\vrule width\thinsize}}

\def\Vb{
   \nextcolumn{\vrule width\thicksize}}


     
{\catcode`\|=13 \let|0
 \catcode`\&=13 \let&0
 \gdef\TableActive{\let|=\vb \let&=\tab}%
}


\def\crrule{\relax                      
   \tstrut                              
   \plaincr\tablerule                   
  }%

\def\crthick{\relax                     
   \tstrut                              
   \plaincr\thickrule                   
  }%
     
\def\crnorule{\relax                    
   \tstrut                              
   \plaincr                             
   }%
   

     
\def\tablerule{\noalign{\hrule height\thinsize depth 0pt}}%
\def\thickrule{\noalign{\hrule height\thicksize depth 0pt}}%


%
%
%
     

\def\linecount{\relax\global\ncols=\curcol      
   \global\curcol=1                             
   \global\advance\nrows by 1\relax}
     
\def\colcount{\relax                            %
   \global\advance\curcol by 1\relax}


\newdimen\parasize      \parasize=4in           

%

%

\def\begintable{\relax                          
    \@BeginRuledTable                           
    \@begintable}

\long\def\@begintable#1\endtable{
   \@RuledTable#1\endruledtable}


\catcode`@=12                                   


\catcode`\@11\relax
\newif\ify@autoscale \y@autoscaletrue \def\Yautoscale#1{\ifnum #1=0
  \y@autoscalefalse\else\y@autoscaletrue\fi}
\newdimen\y@b@xdim
\newdimen\y@boxdim \y@boxdim=13pt
\def\Yboxdim#1{\y@autoscalefalse\y@boxdim=#1}
\newdimen\y@linethick    \y@linethick=.3pt
\def\Ylinethick#1{\y@linethick=#1}
\newskip\y@interspace \y@interspace=0ex plus 0.3ex
\def\Yinterspace#1{\y@interspace=#1}
\newif\ify@vcenter   \y@vcenterfalse
\def\Yvcentermath#1{\ifnum #1=0 \y@vcenterfalse\else\y@vcentertrue\fi}
\newif\ify@stdtext   \y@stdtextfalse
\def\Ystdtext#1{\ifnum #1=0 \y@stdtextfalse\else\y@stdtexttrue\fi}
\newif\ify@enable@skew   \y@enable@skewfalse
\expandafter\ifx\csname enableskew\endcsname\relax
 \y@enable@skewfalse \else \y@enable@skewtrue\fi
\def\y@vr{\vrule height0.8\y@b@xdim width\y@linethick depth 0.2\y@b@xdim}
\def\y@emptybox{\y@vr\hbox to \y@b@xdim{\hfil}}
\ify@enable@skew
 \def\y@abcbox#1{\if :#1\else
   \y@vr\hbox to \y@b@xdim{\hfil#1\hfil}\fi}
 \def\y@mathabcbox#1{\if :#1\else
   \y@vr\hbox to \y@b@xdim{\hfil$#1$\hfil}\fi}
\else
 \def\y@abcbox#1{\y@vr\hbox to \y@b@xdim{\hfil#1\hfil}}
 \def\y@mathabcbox#1{\y@vr\hbox to \y@b@xdim{\hfil$#1$\hfil}}
\fi
\def\y@setdim{%
  \ify@autoscale%
   \ifvoid1\else\typeout{Package youngtab: box1 not free! Expect an
     error!}\fi%
   \setbox1=\hbox{A}\y@b@xdim=1.6\ht1 \setbox1=\hbox{}\box1%
  \else\y@b@xdim=\y@boxdim \advance\y@b@xdim by -2\y@linethick
  \fi}
\newcount\y@counter
\newif\ify@islastarg
\def\y@lastargtest#1,#2 {\if\space #2 \y@islastargtrue
  \else\y@islastargfalse\fi}
\def\y@emptyboxes#1{\y@counter=#1\loop\ifnum\y@counter>0
  \advance\y@counter by -1 \y@emptybox\repeat}
\def\y@nelineemptyboxes#1{%
  \vbox{%
    \hrule height\y@linethick%
    \hbox{\y@emptyboxes{#1}\y@vr}
    \hrule height\y@linethick}\vskip-\y@linethick}
\def\yng(#1){%
  \y@setdim%
  \hskip\y@interspace%
  \ifmmode\ify@vcenter\vcenter\fi\fi{%
  \y@lastargtest#1,
  \vbox{\offinterlineskip
    \ify@islastarg
     \y@nelineemptyboxes{#1}
    \else
     \y@ungempty(#1)
    \fi}}\hskip\y@interspace}
\def\y@ungempty(#1,#2){%
  \y@nelineemptyboxes{#1}
  \y@lastargtest#2,
  \ify@islastarg
   \y@nelineemptyboxes{#2}
  \else
   \y@ungempty(#2)
  \fi}
\def\y@nelettertest#1#2. {\if\space #2 \y@islastargtrue
  \else\y@islastargfalse\fi}
\def\y@abcboxes#1#2.{%
  \ify@stdtext\y@abcbox#1\else\y@mathabcbox#1\fi%
  \y@nelettertest #2.
  \ify@islastarg\unskip%
   \ify@stdtext\y@abcbox{#2}\else\y@mathabcbox{#2}\fi%
  \else\y@abcboxes#2.\fi}
 \newdimen\y@full@b@xdim
 \newcount\y@m@veright@cnt
\ify@enable@skew
 \def\y@get@m@veright@cnt#1#2.{%
   \if :#1 \advance\y@m@veright@cnt by 1\y@get@m@veright@cnt#2.\fi}
 \let\y@setdim@=\y@setdim
 \def\y@setdim{%
   \y@setdim@ \y@full@b@xdim=\y@b@xdim
   \advance\y@full@b@xdim by 1\y@linethick}
 \def\y@m@veright@ifskew#1{
   \y@m@veright@cnt=0 \y@get@m@veright@cnt#1.
   \moveright \y@m@veright@cnt\y@full@b@xdim}
\else
 \def\y@m@veright@ifskew#1{}
\fi
\def\y@nelineabcboxes#1{%
  \y@nelettertest #1.
  \ify@islastarg
   \y@m@veright@ifskew{#1}
    \vbox{
      \hrule height\y@linethick%
      \hbox{\ify@stdtext\y@abcbox#1\else\y@mathabcbox#1\fi\y@vr}
      \hrule height\y@linethick}\vskip-\y@linethick
  \else
   \y@m@veright@ifskew{#1}
    \vbox{
      \hrule height\y@linethick%
      \hbox{\y@abcboxes #1.\y@vr}%
      \hrule height\y@linethick}\vskip-\y@linethick
  \fi}
\def\young(#1){%
  \y@setdim%
  \hskip\y@interspace%
  \y@lastargtest#1,
  \ifmmode\ify@vcenter\vcenter\fi\fi{%
  \vbox{\offinterlineskip
    \ify@islastarg\y@nelineabcboxes{#1}%
    \else\y@ungabc(#1)%
    \fi}}\hskip\y@interspace}
\def\y@ungabc(#1,#2){%
  \y@nelineabcboxes{#1}%
  \y@lastargtest#2,
  \ify@islastarg\y@nelineabcboxes{#2}%
  \else\y@ungabc(#2)%
  \fi}
\catcode`\@12\relax
 



\newfam\scrfam
\batchmode\font\tenscr=rsfs10 \errorstopmode
\ifx\tenscr\nullfont
        \message{rsfs script font not available. Replacing with calligraphic.}
        \def\scr{\cal}
\else   
        \font\sevenscr=rsfs7
        \font\fivescr=rsfs5
        \skewchar\tenscr='177 \skewchar\sevenscr='177 \skewchar\fivescr='177
        \textfont\scrfam=\tenscr \scriptfont\scrfam=\sevenscr
        \scriptscriptfont\scrfam=\fivescr
        \def\scr{\fam\scrfam}
        \def\cal{\scr}
\fi
\catcode`\@=11
\newfam\frakfam
\batchmode\font\tenfrak=eufm10 \errorstopmode
\ifx\tenfrak\nullfont
        \message{eufm font not available. Replacing with italic.}
        
\else
	
	\font\sevenfrak=eufm7 \font\fivefrak=eufm5
	\textfont\frakfam=\tenfrak
	\scriptfont\frakfam=\sevenfrak \scriptscriptfont\frakfam=\fivefrak
	
\fi
\catcode`\@=\active
\newfam\msbfam
\batchmode\font\twelvemsb=msbm10 scaled\magstep1 \errorstopmode
\ifx\twelvemsb\nullfont\def\Bbb{\bf}

	\message{Blackboard bold not available. Replacing with boldface.}
\else   \catcode`\@=11
        \font\tenmsb=msbm10 \font\sevenmsb=msbm7 \font\fivemsb=msbm5
        \textfont\msbfam=\tenmsb
        \scriptfont\msbfam=\sevenmsb \scriptscriptfont\msbfam=\fivemsb
        \def\Bbb{\relax\expandafter\Bbb@}
        \def\Bbb@#1{{\Bbb@@{#1}}}
        \def\Bbb@@#1{\fam\msbfam\relax#1}
        \catcode`\@=\active

\fi
        \font\eightrm=cmr8              \def\xrm{\eightrm}
        \font\eightbf=cmbx8             \def\xbf{\eightbf}
        \font\eightit=cmti10 at 8pt     \def\xit{\eightit}
                
                     
        \font\eightcp=cmcsc8
        \font\eighti=cmmi8              \def\xold{\eighti}
        \font\eightib=cmmib8             \def\xbold{\eightib}
        \font\teni=cmmi10               \def\old{\teni}
        \font\tencp=cmcsc10

        \font\twelvecp=cmcsc10 scaled\magstep1

        \font\sixrm=cmr6

	 at10pt	
	\font\twelvehelvbold=phvb at12pt
	 at14pt
	\font\sixteenhelvbold=phvb at16pt

\def\noblackbox{\overfullrule=0pt}
\noblackbox

\newtoks\headtext
\headline={\ifnum\pageno=1\hfill\else
	\ifodd\pageno{\eightcp\the\headtext}{ }\dotfill{ }{\old\folio}
	\else{\old\folio}{ }\dotfill{ }{\eightcp\the\headtext}\fi
	\fi}
\def\makeheadline{\vbox to 0pt{\vss\noindent\the\headline\break
\hbox to\hsize{\hfill}}
        \vskip2\baselineskip}
\newcount\infootnote
\infootnote=0
\def\foot#1#2{\infootnote=1
\footnote{${}^{#1}$}{\vtop{\baselineskip=.75\baselineskip
\advance\hsize by -\parindent\noindent{\xrm #2}}}\infootnote=0$\,$}
\newcount\refcount
\refcount=1
\newwrite\refwrite
\def\oldsize{\ifnum\infootnote=1\xold\else\old\fi}
\def\ref#1#2{
	\def#1{{{\oldsize\the\refcount}}\ifnum\the\refcount=1\immediate\openout\refwrite=\jobname.refs\fi\immediate\write\refwrite{\item{[{\xold\the\refcount}]} 
	#2\hfill\par\vskip-2pt}\xdef#1{{\noexpand\oldsize\the\refcount}}\global\advance\refcount by 1}
	}
\def\refout{\catcode`\@=11
        \xrm\immediate\closeout\refwrite
        \vskip2\baselineskip
        {\noindent\twelvecp References}\hfill\nobreak\vskip\baselineskip\nobreak
        \baselineskip=.75\baselineskip
        \input\jobname.refs
        \baselineskip=4\baselineskip \divide\baselineskip by 3
        \catcode`\@=\active\rm}
\def\skipref#1{\hbox to15pt{\phantom{#1}\hfill}\hskip-15pt}

\def\hepth#1{\href{http://xxx.lanl.gov/abs/hep-th/#1}{arXiv:hep-th/{\xold#1}}}

\def\arxiv#1#2{\href{http://arxiv.org/abs/#1.#2}{arXiv:{\xold#1}.{\xold#2}}}
\def\jhep#1#2#3#4{\href{http://jhep.sissa.it/stdsearch?paper=#2\%28#3\%29#4}{J. High Energy Phys. {\xbold #1#2} ({\xold#3}) {\xold#4}}}

\def\ATMP#1#2#3{Adv. Theor. Math. Phys. {\xbold#1} ({\xold#2}) {\xold#3}}

\def\CQG#1#2#3{Class. Quantum Grav. {\xbold#1} ({\xold#2}) {\xold#3}}

\def\JHEP{\jhep}

\def\LMP#1#2#3{Lett. Math. Phys. {\xbold#1} ({\xold#2}) {\xold#3}}
\def\MPLA#1#2#3{Mod. Phys. Lett. {\xbf A}{\xbold#1} ({\xold#2}) {\xold#3}}

\def\NPB#1#2#3{Nucl. Phys. {\xbf B}{\xbold#1} ({\xold#2}) {\xold#3}}

\def\PLB#1#2#3{Phys. Lett. {\xbf B}{\xbold#1} ({\xold#2}) {\xold#3}}

\def\PRD#1#2#3{Phys. Rev. {\xbf D}{\xbold#1} ({\xold#2}) {\xold#3}}
\def\PRL#1#2#3{Phys. Rev. Lett. {\xbold#1} ({\xold#2}) {\xold#3}}

\newcount\sectioncount
\sectioncount=0
\def\section#1#2{\global\eqcount=0
	\global\subsectioncount=0
        \global\advance\sectioncount by 1
	\ifnum\sectioncount>1
	        \vskip2\baselineskip
	\fi
\line{\twelvecp\the\sectioncount. #2\hfill}
       \nobreak\vskip.5\baselineskip\nobreak\noindent
        \xdef#1{{\old\the\sectioncount}}}
\newcount\subsectioncount
\def\subsection#1#2{\global\advance\subsectioncount by 1
\vskip.75\baselineskip\noindent\line{\tencp\the\sectioncount.\the\subsectioncount. #2\hfill}\nobreak\vskip.4\baselineskip\nobreak\noindent\xdef#1{{\old\the\sectioncount}.{\old\the\subsectioncount}}}
\def\immediatesubsection#1#2{\global\advance\subsectioncount by 1
\vskip-\baselineskip\noindent\line{\tencp\the\sectioncount.\the\subsectioncount. #2\hfill}\nobreak\vskip.4\baselineskip\nobreak\noindent\xdef#1{{\old\the\sectioncount}.{\old\the\subsectioncount}}}
\newcount\appendixcount
\appendixcount=0
\def\appendix#1{\global\eqcount=0
        \global\advance\appendixcount by 1
        \vskip2\baselineskip\noindent
        \ifnum\the\appendixcount=1
        \hbox{\twelvecp Appendix A: #1\hfill}\nobreak\vskip\baselineskip\nobreak\noindent\fi
    \ifnum\the\appendixcount=2
        \hbox{\twelvecp Appendix B: #1\hfill}\nobreak\vskip\baselineskip\nobreak\noindent\fi
    \ifnum\the\appendixcount=3
        \hbox{\twelvecp Appendix C: #1\hfill}\nobreak\vskip\baselineskip\nobreak\noindent\fi}

\newcount\eqcount
\eqcount=0
\def\Eqn#1{\global\advance\eqcount by 1
\ifnum\the\sectioncount=0
	\xdef#1{{\old\the\eqcount}}
	\eqno({\oldstyle\the\eqcount})
\else
        \ifnum\the\appendixcount=0
	        \xdef#1{{\old\the\sectioncount}.{\old\the\eqcount}}
                \eqno({\oldstyle\the\sectioncount}.{\oldstyle\the\eqcount})\fi
        \ifnum\the\appendixcount=1
	        \xdef#1{{\oldstyle A}.{\old\the\eqcount}}
                \eqno({\oldstyle A}.{\oldstyle\the\eqcount})\fi
        \ifnum\the\appendixcount=2
	        \xdef#1{{\oldstyle B}.{\old\the\eqcount}}
                \eqno({\oldstyle B}.{\oldstyle\the\eqcount})\fi
        \ifnum\the\appendixcount=3
	        \xdef#1{{\oldstyle C}.{\old\the\eqcount}}
                \eqno({\oldstyle C}.{\oldstyle\the\eqcount})\fi
\fi}
\def\eqn{\global\advance\eqcount by 1
\ifnum\the\sectioncount=0
	\eqno({\oldstyle\the\eqcount})
\else
        \ifnum\the\appendixcount=0
                \eqno({\oldstyle\the\sectioncount}.{\oldstyle\the\eqcount})\fi
        \ifnum\the\appendixcount=1
                \eqno({\oldstyle A}.{\oldstyle\the\eqcount})\fi
        \ifnum\the\appendixcount=2
                \eqno({\oldstyle B}.{\oldstyle\the\eqcount})\fi
        \ifnum\the\appendixcount=3
                \eqno({\oldstyle C}.{\oldstyle\the\eqcount})\fi
\fi}
\def\multi{\global\advance\eqcount by 1}
\def\multieq#1#2{\xdef#1{{\old\the\eqcount#2}}
        \eqno{({\oldstyle\the\eqcount#2})}}
\newtoks\url
\def\Href#1#2{\catcode`\#=12\url={#1}\catcode`\#=\active#2}
\def\href#1#2{{#2}}

\parskip=3.5pt plus .3pt minus .3pt
\baselineskip=14pt plus .1pt minus .05pt
\lineskip=.5pt plus .05pt minus .05pt
\lineskiplimit=.5pt
\abovedisplayskip=18pt plus 4pt minus 2pt
\belowdisplayskip=\abovedisplayskip
\hsize=14cm
\vsize=19cm
\hoffset=1.5cm
\voffset=1.8cm
\frenchspacing
\footline={}
\raggedbottom

\def\ss{\scriptstyle}
\def\sss{\scriptscriptstyle}
\def\*{\partial}
\def\punkt{\,\,.}
\def\komma{\,\,,}

\def\={\!=\!}
\def\small#1{{\hbox{$#1$}}}

\def\fraction#1{\small{1\over#1}}
\def\fr{\fraction}
\def\Fraction#1#2{\small{#1\over#2}}
\def\Fr{\Fraction}
\def\tr{\hbox{\rm tr}}
\def\eg{{\tenit e.g.}}

\def\ie{{\tenit i.e.}}

\def\nlni{\hfill\break}

\def\a{\alpha}
\def\b{\beta}

\def\d{\delta}

\def\g{\gamma}
\def\l{\lambda}

\def\ra{\rightarrow}

\def\ra{\rightarrow}
\def\la{\leftarrow}

\def\rarrowover#1{\vtop{\baselineskip=0pt\lineskip=0pt
      \ialign{\hfill##\hfill\cr$\ra$\cr$#1$\cr}}}

\def\larrowover#1{\vtop{\baselineskip=0pt\lineskip=0pt
      \ialign{\hfill##\hfill\cr$\la$\cr$#1$\cr}}}


\def\modprod#1{\raise0pt\vtop{\baselineskip=0pt\lineskip=0pt
      \ialign{\hfill##\hfill\cr$\circ$\cr${\sss #1}$\cr}}}

\def\eqskip{\qquad}


\ref\BaggerLambertI{J. Bagger and N. Lambert, {\xit ``Modeling
multiple M2's''}, \PRD{75}{2007}{045020} [\hepth{0611108}].}

\ref\BaggerLambertII{J. Bagger and N. Lambert, {\xit ``Gauge symmetry
and supersymmetry of multiple M2-branes''}, \PRD{77}{2008}{065008}
[\arxiv{0711}{0955}].} 

\ref\Gustavsson{A. Gustavsson, {\xit ``Algebraic structures on
parallel M2-branes''}, \arxiv{0709}{1260}.}

\ref\CederwallNilssonTsimpisI{M. Cederwall, B.E.W. Nilsson and D. Tsimpis,
{\xit ``The structure of maximally supersymmetric super-Yang--Mills
theory --- constraining higher order corrections''},
\jhep{01}{06}{2001}{034} 
[\hepth{0102009}].}

\ref\CederwallNilssonTsimpisII{M. Cederwall, B.E.W. Nilsson and D. Tsimpis,
{\xit ``D=10 super-Yang--Mills at $\ss O(\a'^2)$''},
\JHEP{01}{07}{2001}{042} [\hepth{0104236}].}

\ref\BerkovitsParticle{N. Berkovits, {\xit ``Covariant quantization of
the superparticle using pure spinors''}, \jhep{01}{09}{2001}{016}
[\hepth{0105050}].}

\ref\SpinorialCohomology{M. Cederwall, B.E.W. Nilsson and D. Tsimpis,
{\xit ``Spinorial cohomology and maximally supersymmetric theories''},
\jhep{02}{02}{2002}{009} [\hepth{0110069}];
M. Cederwall, {\xit ``Superspace methods in string theory, supergravity and gauge theory''}, Lectures at the XXXVII Winter School in Theoretical Physics ``New Developments in Fundamental Interactions Theories'',  Karpacz, Poland,  Feb. 6-15, 2001, \hepth{0105176}.}

\ref\Movshev{M. Movshev and A. Schwarz, {\xit ``On maximally
supersymmetric Yang--Mills theories''}, \NPB{681}{2004}{324}
[\hepth{0311132}].}

\ref\BerkovitsI{N. Berkovits,
{\xit ``Super-Poincar\'e covariant quantization of the superstring''},
\jhep{00}{04}{2000}{018} [\hepth{0001035}].}

\ref\BerkovitsNonMinimal{N. Berkovits,
{\xit ``Pure spinor formalism as an N=2 topological string''},
\jhep{05}{10}{2005}{089} [\hepth{0509120}].}

\ref\CederwallNilssonSix{M. Cederwall and B.E.W. Nilsson, {\xit ``Pure
spinors and D=6 super-Yang--Mills''}, \arxiv{0801}{1428}.}

\ref\CGNN{M. Cederwall, U. Gran, M. Nielsen and B.E.W. Nilsson,
{\xit ``Manifestly supersymmetric M-theory''},
\JHEP{00}{10}{2000}{041} [\hepth{0007035}];
{\xit ``Generalised 11-dimensional supergravity''}, \hepth{0010042}.
}

\ref\CGNT{M. Cederwall, U. Gran, B.E.W. Nilsson and D. Tsimpis,
{\xit ``Supersymmetric corrections to eleven-dimen\-sional supergravity''},
\jhep{05}{05}{2005}{052} [\hepth{0409107}].}

\ref\NilssonPure{B.E.W.~Nilsson,
{\xit ``Pure spinors as auxiliary fields in the ten-dimensional
supersymmetric Yang--Mills theory''},
\CQG3{1986}{{\xrm L}41}.}

\ref\HowePureI{P.S. Howe, {\xit ``Pure spinor lines in superspace and
ten-dimensional supersymmetric theories''}, \PLB{258}{1991}{141}.}

\ref\HowePureII{P.S. Howe, {\xit ``Pure spinors, function superspaces
and supergravity theories in ten and eleven dimensions''},
\PLB{273}{1991}{90}.} 



\ref\CederwallThreeConf{M. Cederwall, {\xit ``N=8 superfield formulation of
the Bagger--Lambert--Gustavsson model''}, \jhep{08}{09}{2008}{116}
[\arxiv{0808}{3242}]; {\xit ``Superfield actions for N=8 
and N=6 conformal theories in three dimensions''},
\jhep{08}{10}{2008}{70}
[\arxiv{0808}{3242}]; {\xit ``Pure spinor superfields,
with application to D=3 conformal models''}, \arxiv{0906}{5490}.}

\ref\MarneliusOgren{R. Marnelius and M. \"Ogren, {\xit ``Symmetric
inner products for physical states in BRST quantization''},
\NPB{351}{1991}{474}.} 

\ref\BerkovitsICTP{N. Berkovits, {\xit ``ICTP lectures on covariant
quantization of the superstring''}, proceedings of the ICTP Spring
School on Superstrings and Related Matters, Trieste, Italy, 2002
[\hepth{0209059}.]} 

\ref\ABJM{O. Aharony, O. Bergman, D.L. Jafferis and J. Maldacena,
{\xit ``N=6 superconformal Chern--Simons-matter theories, M2-branes
and their gravity duals''}, \arxiv{0806}{1218}.}

\ref\ElevenSG{E. Cremmer, B. Julia and J. Sherk, 
{\xit ``Supergravity theory in eleven-dimensions''},
\PLB{76}{1978}{409}.}

\ref\ElevenSGSuperspace{L. Brink and P. Howe, 
{\xit ``Eleven-dimensional supergravity on the mass-shell in superspace''},
\PLB{91}{1980}{384};
E. Cremmer and S. Ferrara,
{\xit ``Formulation of eleven-dimensional supergravity in superspace''},
\PLB{91}{1980}{61}.}
 
\ref\BatalinVilkovisky{I.A. Batalin and G.I. Vilkovisky, {\xit ``Gauge
algebra and quantization''}, \PLB{102}{1981}{27}.}

\ref\FusterBVReview{A. Fuster, M. Henneaux and A. Maas, {\xit
``BRST-antifield quantization: a short review''},\nlni\hepth{0506098}.}

\ref\CederwallInProgress{M. Cederwall, work in progress.}

\ref\ZwiebachClosedBV{B. Zwiebach, {\xit ``Closed string field theory:
    Quantum action and the BV master equation''}, \hepth{9206084}.}

\ref\BerkovitsMembrane{N. Berkovits,
	{\xit ``Towards covariant quantization of the supermembrane''},
	\JHEP{02}{09}{2002}{051} [\hepth{0201151}].}

\ref\BerkovitsNekrasovCharacter{N. Berkovits and N. Nekrasov, {\xit
    ``The character of pure spinors''}, \LMP{74}{2005}{75}
  [\hepth{0503075}].}

\ref\BerkovitsNekrasovMultiloop{N. Berkovits and N. Nekrasov, {\xit
    ``Multiloop superstring amplitudes from non-minimal pure spinor
    formalism''}, \jhep{06}{12}{2006}{029} [\hepth{0609012}].}

\ref\HoweWeyl{P. Howe, {\xit ``Weyl superspace''},
  \PLB{415}{1997}{149} [\hepth{9707184}].}

\ref\BoulangerUniqueness{N. Boulanger, T. Damour, L. Gualtieri and
  M. Henneaux, {\xit ``Inconsistency of interacting, multigraviton
    theories''}, \NPB{597}{2001}{127} [\hepth{0007220}].}

\ref\GAMMA{U. Gran,
{\xit ``GAMMA: A Mathematica package for performing gamma-matrix 
algebra and Fierz transformations in arbitrary dimensions''},
\hepth{0105086}.}

\ref\AnguelovaGrassiVanhove{L. Anguelova, P.A. Grassi and P. Vanhove,
  {\xit ``Covariant one-loop amplitudes in D=11''},
  \NPB{702}{2004}{269} [\hepth{0408171}].}

\ref\GrassiVanhove{P.A. Grassi and P. Vanhove, {\xit ``Topological M
    theory from pure spinor formalism''}, \ATMP{9}{2005}{285}
  [\hepth{0411167}].} 

\ref\PureSG{M. Cederwall, {\xit ``Towards a manifestly supersymmetric
    action for D=11 supergravity''}, \jhep{10}{01}{2010}{117}
    [\arxiv{0912}{1814}],  
{\xit ``D=11 supergravity with manifest supersymmetry''},
    \MPLA{25}{2010}{3201} [\arxiv{1001}{0112}].}

\ref\AndreevTseytlinBIDerCorr{O.D. Andreev and A.A. Tseytlin,
{\xit ``Partition function representation for the open superstring 
effective action: cancellation of Mobius infinities and derivative 
corrections to Born--Infeld lagrangian''},
\NPB{311}{1988}{205}.}

\ref\BergshoeffKAPPA{E.A.~Bergshoeff, M.~de~Roo and A.~Sevrin,
{\xit On the supersymmetric non-abelian Born--Infeld action''},
\hepth{0011264}.}

\ref\BergshoeffFFOUR{E.~Bergshoeff, M.~Rakowski and E.~Sezgin,
{\xit ``Higher derivative super Yang--Mills theories},
\PLB{185}{1987}{371}.}

\ref\BornInfeldLagr{E.S. Fradkin and A.A. Tseytlin, 
{\xit ``Non-linear electrodynamics from quantized strings''},
\PLB{163}{1985}{123}.}

\ref\Leigh{R.G. Leigh, {\xit ``Dirac--Born--Infeld action from Dirichlet sigma
	model''}, Mod. Phys. Lett. {\xbf A4} ({\xold1989}) {\xold2767};\nlni
	C.G. Callan, C. Lovelace, C.R. Nappi and S.A. Yost,
	{\xit ``String loop corrections to beta functions''},
	Nucl. Phys. {\xbf B288} ({\xold1987}) {\xold525}.}

\ref\SevrinFSIX{A. Sevrin, J. Troost and W. Troost,
{\xit ``The non-abelian Born--Infeld action at order $\ss F^6$''},
\hepth{0101192}.}

\ref\TseytlinBIREV{A.A.~Tseytlin, 
{\xit Born--Infeld action, supersymmetry and string theory},
\xrm in the Yuri Golfand memorial volume, ed. M. Shifman,
World Scientific (2000) [\hepth{9908105}]}

\ref\TseytlinSTR{A.A.~Tseytlin, 
{\xit "On the non-abelian generalization of 
Born--Infeld action in string theory"},
\NPB{501}{1997}{41} [\hepth{9701125}].}

\ref\BornInfeld{M. Born and L. Infeld, {\xit ``Foundations of the new
field theory''}, Proc. Royal Soc. London {\xbf A144} (1934) 425.}

\ref\DBranesI{M. Cederwall, A. von Gussich, B.E.W. Nilsson and A. Westerberg,
{\xit ``The Dirichlet super-three-brane in ten-dimensional type IIB 
supergravity''}
\NPB{490}{1997}{163} [\hepth{9610148}].}

\ref\Aganagic{M. Aganagi\'c, C. Popescu, J.H. Schwarz,
{\xit ``D-brane actions with local kappa symmetry''},
\PLB{393}{1997}{311} [\hepth{9610249}].}

\ref\DBranesII{M. Cederwall, A. von Gussich, B.E.W. Nilsson, P. Sundell
 and A. Westerberg,
{\xit ``The Dirichlet super-p-branes in ten-dimensional type IIA and IIB 
supergravity''},
\NPB{490}{1997}{179} [\hepth{9611159}].}

\ref\BergshoeffTownsendDBranes{E. Bergshoeff and P.K. Townsend, 
{\xit ``Super D-branes''},
\NPB{490}{1997}{145}\nlni[\hepth{9611173}].}

\ref\Polchinski{J. Polchinski,
{\xit ``Dirichlet branes and Ramond-Ramond charges''},
\PRL{75}{1995}{4724}\nlni[\hepth{9510017}].} 

\ref\BerkovitsPershinBI{N. Berkovits and V. Pershin, {\xit
    ``Supersymmetric Born-Infeld from the pure spinor formalism of the
    open superstring''}, \JHEP{03}{01}{2003}{023} [\hepth{0205154}].}

\ref\GibbonsBI{G.W. Gibbons, {\xit ``Aspects of Born-Infeld theory and
string/M-theory''}, Rev. Mex. Fis. {\xbf 49S1} (2003) 19 [\hepth{0106059}].}

\ref\BergshoeffBIRevisited{E. Bergshoeff, A. Bilal, M. de Roo and
A. Sevrin, {\xit ``Supersymmetric non-abelian Born-Infeld
revisited''}, \JHEP{01}{07}{2001}{029} [\hepth{0105274}].}

\ref\MetsaevBI{R.R. Metsaev, M. Rakhmanov and A.A. Tseytlin, {\xit
``The Born-Infeld action as the effective action in the open
superstring theory''}, \PLB{193}{1987}{207}.}

\ref\CecottiFerrara{S. Cecotti and S. Ferrara, {\xit ``Supersymmetric
Born-Infeld Lagrangians''}, \PLB{187}{1987}{335}.}

\ref\SuperYM{L. Brink, J.H. Schwarz and J. Scherk, 
{\xit ``Supersymmetric Yang--Mills theories''},
\NPB{121}{1977}{77}.}

\ref\NilssonSYM{B.E.W.~Nilsson, 
\xit ``Off-shell fields for the 10-dimensional supersymmetric 
Yang--Mills theory'', \xrm G\"oteborg-ITP-{\xold81}-{\xold6}.}

\ref\MovshevDef{M. Movshev, {\xit ``Deformation of maximally
supersymmetric Yang--Mills theory in dimensions 10. An algebraic
approach''},
\hepth{0601010}.}

\ref\MovshevSchwarzDef{M. Movshev and A. Schwarz, {\xit
``Supersymmetric deformations of maximally supersymmetric gauge
theories. I''}, \arxiv{0910}{0620}.}

\ref\TsimpisPolicastro{G. Policastro and D. Tsimpis, 
{\xit ``$\ss R^4$, purified''}, [\hepth{0603165}].}

\ref\HoweLindstromWulff{P.S. Howe, U. Lindstr\"om and L. Wulff, 
{\xit ``D=10 supersymmetric Yang-Mills theory at
$\ss O(\a'^4)$''}, \jhep{10}{07}{2010}{028} [\arxiv{1004}{3466}].}

\ref\BHLSW{G. Bossard, P.S. Howe, U. Lindstr\"om, K.S. Stelle and
L. Wulff, {\xit ``Integral invariants in maximally supersymmetric
Yang-Mills theories''}, \arxiv{1012}{3142}.}

\ref\BerkovitsNekrasovMultiloop{N. Berkovits and N. Nekrasov, {\xit
    ``Multiloop superstring amplitudes from non-minimal pure spinor
    formalism''}, \jhep{06}{12}{2006}{029} [\hepth{0609012}].}

\ref\BerkovitsII{N. Berkovits and B.C. Valillo, 
{\xit ``Consistency of super-Poincar\'e covariant superstring tree
amplitudes''}, \jhep{00}{07}{2000}{015} [\hepth{0004171}].}

\ref\BjornssonGreen{J. Bj\"ornsson and M.B. Green, {\xit ``5 loops in
25/4 dimensions''}, \jhep{10}{08}{2010}{132} [\arxiv{1004}{2692}].}

\ref\BjornssonMultiLoop{J. Bj\"ornsson, {\xit ``Multi-loop amplitudes
in maximally supersymmetric pure spinor field
theory''}, \jhep{11}{01}{2011}{002} [\arxiv{1009}{5906}].}

\ref\Bossard{G. Bossard, P.S. Howe, K.S. Stelle and P. Vanhove, {\xit
``The vanishing volume of D=4 superspace''},\nlni\arxiv{1105}{6087}.}

\ref\GrassiVanhoveAmpl{P.A. Grassi and P. Vanhove, {\xit ``Higher-loop
amplitudes in the non-minimal pure spinor
formalism''}, \jhep{09}{05}{2009}{089} [\arxiv{0903}{3903}].}

\ref\BerkovitsGreenRussoVanhove{N. Berkovits, M.B. Green, J. Russo and
    P. Vanhove, {\xit ``Non-renormalization conditions for four-gluon
    scattering in supersymmetric string and field
    theory''}, \jhep{09}{11}{2009}{063} [\arxiv{0908}{1923}].}

\ref\HoweTsimpis{P.S. Howe and D. Tsimpis, {\xit ``On higher-order
corrections in M theory''}, \jhep{03}{09}{2003}{038}
[\hepth{0305129}].}

\ref\WittenDBranes{E. Witten, 
{\xit ``Bound states of strings and p-branes''},
\NPB{460}{1996}{335} [\hepth{9510135}].}

\ref\HarmonicSuperspace{A.S. Galperin, E.A. Ivanov, V.I. Ogievetsky
and E.S. Sokatchev, {\xit ``Harmonic superspace''}, Cambridge
University Press (2001).}

\ref\LiE{A.M. Cohen, M. van Leeuwen and B. Lisser, 
LiE v. {\xold2}.{\xold2} ({\xold1998}), 
\nlni http://wallis.univ-poitiers.fr/\~{}maavl/LiE/} 

\ref\GAMMA{U. Gran,
{\xit ``GAMMA: A Mathematica package for performing gamma-matrix 
algebra and Fierz transformations in arbitrary dimensions''},
\hepth{0105086}.}

\ref\CollinucciDeRooEeninkYM{A. Collinucci, M. de Roo and M.G.C. Eenink,
{\xit ``Supersymmetric Yang--Mills theory at order
$\ss \a'^3$''}, \jhep{02}{06}{2002}{024} [\hepth{0205150}].}

\ref\CollinucciDeRooEeninkMaxwell{A. Collinucci, M. de Roo and M.G.C. Eenink,
{\xit ``Derivative corrections in ten-dimensional super-Maxwell
theory''}, \jhep{03}{01}{2003}{039} [\hepth{0212012}].} 

\ref\CederwallNilssonTsimpisIII{M. Cederwall, B.E.W. Nilsson and D. Tsimpis,
{\xit ``Spinorial cohomology of abelian D=10 super-Yang--Mills at $\ss
O(\a'^3)$''}, 
\JHEP{02}{11}{2002}{023} [\hepth{0205165}].}

\ref\BeisertEtAlSG{N. Beisert, H. Elvang, D.Z. Freedman, M. Kiermaier,
A. Morales and S. Stieberger, {\xit ``E7(7) constraints on
counterterms in N=8 supergravity''}, \PLB{694}{2010}{265} [\arxiv{1009}{1643}].}


\headtext={M. Cederwall, A. Karlsson: 
``Pure spinor superfields and Born--Infeld theory''}

\line{
\epsfxsize=18mm
\epsffile{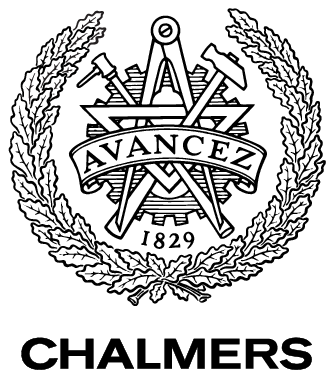}
\hfill}
\vskip-12mm
\line{\hfill Gothenburg preprint}
\line{\hfill September, {\old2011}}
\line{\hrulefill}

\vfill
\vskip.5cm

\centerline{\sixteenhelvbold
Pure spinor superfields and Born--Infeld theory} 

%

\vfill

\centerline{\twelvehelvbold{Martin Cederwall}}
\vskip2\parskip
\centerline{\twelvehelvbold{Anna Karlsson}}

\vfill
\vskip-1cm

\centerline{\xrm Fundamental Physics}
\centerline{\xrm Chalmers University of Technology}
\centerline{\xrm SE 412 96 Gothenburg, Sweden}

\vfill

{\narrower\noindent \underbar{Abstract:} We present a method for
introducing and analysing higher-derivative deformations of maximally
supersymmetric field theories. Such terms are built in the pure spinor
superfield framework, using a set
of operators representing physical fields. 
The action for abelian Born--Infeld theory becomes polynomial in this
language, and 
contains only a four-point interaction in addition to the free
action. Simplifications also occur in the non-abelian case.
\smallskip}
\vfill

\font\xxtt=cmtt6

\vtop{\baselineskip=.6\baselineskip\xxtt
\line{\hrulefill}
\catcode`\@=11
\line{email: martin.cederwall@chalmers.se, karann@chalmers.se\hfill}
\catcode`\@=\active
}

\eject

\def\textfrac#1#2{\raise .45ex\hbox{\the\scriptfont0 #1}\nobreak\hskip-1pt/\hskip-1pt\hbox{\the\scriptfont0 #2}}

\def\lb{\bar\l}

\section\Intro{Introduction}Pure spinors have a long history
in connection to (maximal)
supersymmetry and superstring theory 
[\NilssonPure-\skipref\HowePureI\skipref\HowePureII\skipref\BerkovitsI\skipref\BerkovitsParticle\skipref\CederwallNilssonTsimpisI\skipref\CederwallNilssonTsimpisII\skipref\SpinorialCohomology\Movshev]. The
main achievement of the formalism is that it solves the problem of
manifesting supersymmetry, without breaking other symmetries, in
maximally supersymmetric models. 

Variations of the pure spinor formalism has been successfully used for
calculation of perturbative amplitudes in maximally supersymmetric
gauge theory and superstring theory
[\BerkovitsII-\skipref\BerkovitsNekrasovMultiloop\skipref\GrassiVanhoveAmpl\skipref\BerkovitsGreenRussoVanhove\skipref\BjornssonGreen\BjornssonMultiLoop]. A
partly overlapping activity has been the search for ``supersymmetric
invariants'', \ie, possible counterterms or terms in a low-energy
effective action
[\CederwallNilssonTsimpisI,\CederwallNilssonTsimpisII,\CGNN-\skipref\CGNT\skipref\HoweTsimpis\TsimpisPolicastro]. 

It has become clear during the past few years that the most efficient
use of the pure spinor formalism (in field theory) is when the full
Batalin--Vilkovisky (BV) field--antifield
formalism [\BatalinVilkovisky,\FusterBVReview] is used. The language
is that of classical BV theory. The canonical example is $D=10$
super-Yang--Mills (SYM) theory [\SuperYM], which will be reviewed below,
but also other maximally supersymmetric
field theories, including supergravity, have been formulated in this
framework [\CederwallThreeConf,\PureSG]. The resulting
actions generically are ``simpler'' than the component ones, and
contain terms with lower maximal degree of homogeneity. In certain
cases, an action which is non-polynomial in physical component fields
becomes polynomial [\PureSG]. 
This of course inspires to ask similar
question concerning supersymmetric invariants in general: Can they
show similar simplifications? 

An important example of a supersymmetric invariant, possessing some
remarkable properties, and relevant in string theory as the effective
theory of supersymmetric D-branes 
[\Polchinski-\skipref\DBranesI\skipref\Aganagic\skipref\DBranesII\BergshoeffTownsendDBranes],
is the Born--Infeld theory [\BornInfeld]. 
Much work has been done on Born--Infeld theory, its supersymmetric
versions, and its r\^ole in string theory, see \eg\ refs.  
[\MetsaevBI-\skipref\CecottiFerrara\skipref\AndreevTseytlinBIDerCorr\skipref\Leigh\skipref\BornInfeldLagr\skipref\BergshoeffKAPPA\skipref\BergshoeffFFOUR\skipref\SevrinFSIX\skipref\TseytlinBIREV\skipref\TseytlinSTR\skipref\BergshoeffBIRevisited\GibbonsBI]. 
While the abelian Born--Infeld Lagrangian (and its supersymmetrisation)
is well known, the non-abelian counterpart, describing coincident
D-branes [\WittenDBranes], has been elusive. The full term at order
$\alpha'{}^2$ (the ``$F^4$ term'') was first constructed in
ref. [\CederwallNilssonTsimpisII], and extended to the next order
($\alpha'{}^4$) in ref. [\HoweLindstromWulff]. 
No closed expression
has been proposed, consistent with the known terms to this
order.

One main motivation for the present investigation is to examine
whether Born--Infeld dynamics exhibits any simplification in the pure
spinor superfield formalism. We will see that this is indeed the
case. The abelian Born--Infeld action becomes polynomial, with only an
$\a'{}^2$ term in addition to the kinetic term, and consequently the
$\a'{}^2$ term contains all information on the symmetrised trace part
in the non-abelian situation.

Born--Infeld theory has been considered earlier in connection to pure
spinors in ref. [\BerkovitsPershinBI], where pure spinor superstrings
ending on a D-brane were constructed. Unlike our treatment, where
non-minimal variables are needed, the obtained superspace equations of
motion are contained in a minimal pure spinor setting, and are non-polynomial.

Another motivation is the systematic search for invariants in
general. 
Collinucci, de Roo and Eenink showed in
ref. [\CollinucciDeRooEeninkMaxwell], using a component formalism, 
that there are two independent
quartic deformations of $D=10$ Maxwell theory. In addition to
the one at order $\a'{}^2$ given below, 
there is also one at order $\a'{}^4$. They also
constructed a linear deformation of $D=10$ super-Yang--Mills theory at
order $\a'{}^3$ whose lowest order terms are quartic 
[\CollinucciDeRooEeninkYM]. There is no non-trivial deformation at
$\a'{}^3$ in Maxwell theory 
[\CollinucciDeRooEeninkYM,\CederwallNilssonTsimpisIII]. 
Movshev and Schwarz have given a classification of infinitesimal
deformations of $D=10$ super-Yang--Mills theory and its dimensional
reductions [\MovshevDef,\MovshevSchwarzDef]. We
would like to connect to their results, but will leave this for the future.

Let us sketch our main line of thought. Earlier approaches to
deformations (higher-derivative terms) in the pure spinor formalism
have used only part of the information available, namely the equations
for the physical superfields at ghost number 0. This is logical in a
certain sense --- although the cohomology contains also ghosts and
antifields, the gauge transformations should not be deformed. But if
we want to use the full BV formalism, which means that the master
equation is what defines a consistent deformation, we need to retain
the full pure spinor superfields. Still, deformations are built from
physical fields, and we need a way to ``obtain them'' from the
field. More specifically, we need to find operators of ghost number
$-1$ that act on the pure spinor superfield $\Psi$ and give new
pure spinor superfields of ghost number 0, transforming as physical fields.
We also need to understand the cohomology of such fields.
In refs. [\CederwallNilssonTsimpisI,\CederwallNilssonTsimpisII], the
$\a'{}^2$ terms was encoded in the superspace relation
$$
F^A_{\a\b}\sim\a'{}^2t^A{}_{BCD}(\g^i\chi^B)_\a(\g^j\chi^C)_\b F^D_{ij}
\komma\Eqn\OldDeformation
$$
where $\chi$ is the physical fermion, and $t_{ABCD}$ a totally
symmetric tensor.
The physical fields on the right hand side need to be reexpressed
using operators of negative ghost number.

Interesting results have been obtained using harmonic
superspace (see \eg\ refs. [\Bossard,\HarmonicSuperspace] 
and references therein). 
We will not try to make
any comparison between such approaches and the one advocated in the
present paper.

The paper is organised as follows. In Section {\old2}, the pure spinor
superfield BV formalism for $D=10$ SYM is briefly reviewed.
Section {\old3} deals with non-scalar superfields and their gauge
symmetries. The operators corresponding to physical fields are derived
in Section {\old4}, and used in Section {\old5} to construct
interaction terms respecting the BV master equation. Our results are
summarised in Section {\old6}, where we also propose further lines of
exploration. Some conventions are given in Appendix A. A few useful
spinor identities are listed in Appendix B, 
and Appendix C gives the list of modules appearing in polynomials in
the non-minimal pure spinor variables.  

\section\PureSpinorBV{Batalin--Vilkovisky actions for pure spinor
fields}It is well known that the superspace formulation of $D=10$
super-Yang--Mills theory is contained in a scalar fermionic pure
spinor superfield $\Psi(x,\theta;\l)$ of ghost number 1, subject to
the equation of motion 
$$
Q\Psi+\Psi^2=0\punkt\Eqn\CSEOM
$$
Here, $\l^\a$ is a pure spinor, \ie, $(\l\g^a\l)=0$. The BRST operator
$Q$ is defined as $Q=(\l D)$, $D_\a$ being the fermionic
covariant derivative. 
(We also often suppress a Lie algebra index on the field. For example,
$(\Psi^2)^A=\fr2f^A{}_{BC}\Psi^B\Psi^C$, where $f$ are structure
constants.) 

This is the canonical example of a pure spinor superfield theory. The
other example with a scalar field is $D=11$ supergravity
[\PureSG]. Much of what is said here applies also to that model.

The standard
on-shell superspace description is obtained by restricting to
a superfield of ghost number 0, $\Psi(x,\theta;\l)=\l^\a
A_\a(x,\theta)$. The equation of motion then is the 5-form part of the
superspace flatness condition [\NilssonSYM]
$$
F_{\a\b}\equiv 2D_{(\a} A_{\b)}+\{A_\a,A_\b\}+2\g_{\a\b}^iA_i=0\punkt\Eqn\SSFlatness
$$
The vector part of this equation is the usual conventional constraint.

In addition to describing the physical fields, the solution of 
equation (\CSEOM) gives the ghost field at ghost number 1, and the
antifields of the physical and ghost fields at ghost number $-1$ and
$-2$, respectively. The natural language becomes the field--antifield
formalism of Batalin and Vilkovisky
[\BatalinVilkovisky,\FusterBVReview].

With an appropriately defined integration, the equations of motion
come from a Chern--Simons-like action 
$$
S_{\hbox{\sixrm CS}}=\int[dZ]\tr\left(\fr2\Psi Q\Psi+\fr3\Psi^3\right)\komma
\Eqn\CSAction
$$
and one will then have an supersymmetric off-shell formulation of the
theory. Defining the integration (and regularising integrals) calls for
the introduction of non-minimal pure spinor variables
[\BerkovitsNonMinimal]. In addition to the pure spinor $\l^\a$, a pure
spinor $\lb_\a$ of the opposite chirality is introduced, together with
a fermionic spinor $r_\a$ satisfying $(\lb\g^a r)=0$. The BRST
operator is extended to 
$$
Q=(\l D)+(r{\*\over\*\lb})\komma\eqn
$$
which leaves its cohomology unchanged. We refer to
refs. [\BerkovitsNonMinimal,\BerkovitsNekrasovMultiloop] for details.

How is the consistency of the action (\CSAction) checked? This is
especially relevant when we will try to add higher-derivative
terms. One may check that the action is invariant under gauge
transformations
$\delta\Psi=Q\Lambda+[\Psi,\Lambda]$. But a more
efficient way is to take the Batalin--Vilkovisky formalism {\it ad
notam}. In this framework, the action itself is the generator of
``gauge transformations'' in a generalised sense, via the antibracket.
Invariance of the action itself is encoded in the master equation
$$
(S,S)=0\komma\eqn
$$
which is also the appropriate generalisation of ``$Q^2=0$'' to an
interacting theory. 
The antibracket in pure spinor field theory takes the simplest
possible form [\PureSG]:
$$
(A,B)=\int
A{\larrowover\d\over\d\Psi^A(Z)}[dZ]{\rarrowover\d\over\d\Psi^A(Z)}B
\punkt\eqn
$$
The field $\Psi$ is self-conjugate with respect to the antibracket.

This makes the master equation reasonably easy to check, also when we
later add terms corresponding to higher-derivative corrections. These
terms (seen as infinitesimal deformations) also have a clear
cohomological interpretation: they are additional terms $S'$ in the
action, with the property of being closed, $(S,S')=0$ but counted
modulo exact terms (corresponding to field redefinitions),
$S'\approx S'+(S,R)$.   

\section\Shift{Non-scalar fields and shift symmetries}The basic field
in $D=10$ (or $D=4$, $N=4$) super-Yang--Mills theory is scalar
(because the ghost is a scalar). BRST cohomology arises thanks to the
pure spinor constraint. For example, consider the zero mode of the
gauge connection. It sits in the field as
$\Psi\sim(\l\g^i\theta)A_i$. Acting with $Q$ gives the pure spinor
constraint. It is also clear that no other (zero mode) cohomology than
the ghost cohomology $\Psi\sim1$ exists at $\l^0$. But what about
fields of ghost number zero, describing 
supermultiplets? In some cases (though not in
$D=10$) one needs to describe hypermultiplets, which have no gauge
invariance. It they are to be described by a pure spinor superfield,
which we may call $\Phi^I$,
the field will need to have ghost number 0 and come in the same module
(of Lorentz and R-symmetry) as the scalar component fields. In
addition, one needs the fermionic fields to be represented by
cohomology at $\l^0\theta^1$. It is clear that this cannot be achieved
by simply demanding $Q\Phi^I=0$. This cohomology will simply be the
tensor product of the cohomology of a scalar field with the module of
$\Psi^I$, and there is no room for the fermions. 
In $D=10$, although no non-scalar field is needed to describe the SYM
multiplet, we have argued in the introduction that it will be
necessary to introduce derived fields transforming in the modules of
physical fields. This construction will be performed in the following
Section. So, the situation is similar.
Some other ingredient than the pure spinor constraint is needed.

Suppose that, in addition to the pure spinor constraint, which should
be seen as a gauge symmetry defining the equivalence class
$\Psi\approx\Psi+(\l\g^i\l)\Xi_i$, one has a further gauge symmetry
involving the index structure of the field. Suppose, in the example
with a hypermultiplet above (although the argument is completely
general), that we have scalars $\phi^I$ and
fermions $\chi^{\tilde\a}$. $\chi$ should come at level $\theta$
in a superfield starting with $\phi$, 
$$
\Phi^I(x,\theta)=\phi^I(x)+R^I_{\a\tilde\a}\theta^\a\chi^{\tilde\a}+\ldots
\Eqn\MatterField
$$ 
(the invariant tensor $R$ is typically a $\g$-matrix). 
Now,
we demand that (at least the zero mode of) the fermion arises as pure
spinor BRST cohomology. Acting with $Q$ on the superfield
(\MatterField) gives 
$$
Q\Phi^I\sim R^I_{\a\tilde\a}\l^\a\chi^{\tilde\a}+\ldots\punkt\eqn
$$
The fermions will represent cohomology only if this expression is
``zero'', \ie, if fields are taken in the equivalence class
$$
\Phi^I\approx\Phi^I+R^I_{\a\tilde\a}\l^\a\xi^{\tilde\a}\punkt\eqn
$$
We call this a ``shift symmetry''. It should be thought of on an equal
footing as the pure spinor constraint.

Such constructions have been relevant for the supersymmetric
descriptions of BLG and ABJM models in $D=3$
[\CederwallThreeConf]. It was also essential in $D=11$
supergravity [\PureSG], where the vector field $\Phi^a$ corresponding to the
superspace geometry was constructed as an operator $R^a$ acting on the
fundamental scalar field $\Psi$ corresponding to the tensor field.

As already demonstrated by the supergravity application, 
the principle is not only relevant for describing matter multiplets. 
In the following Section, we will show how to reinterpret the
superspace Bianchi identities in terms of pure spinor superfields of
ghost number 0, which will be obtained from $\Psi$ by acting with
operators of ghost number $-1$ with certain properties interpreted as
shift symmetry.

This will provide the concrete answer to the question posed in
the Introduction how the ghost number 0 superfields can be extended
to pure spinor fields and used to construct deformation terms in the action.

\vfill\eject

\section\Operators{Physical operators}
\immediatesubsection\FromSSToOps{From superspace to operators}Let us 
go back to the superspace
equations of motion (\SSFlatness),  
implied by eq. (\CSEOM), which follow from the action
$$
\eqalign{
S_{\hbox{\sixrm CS}}=S_2+S_3&=\int[dZ]\tr\left(\fr2\Psi Q\Psi+\fr3\Psi^3\right)\cr
&=\int[dZ]\left(\fr2\Psi^AQ\Psi^A+\fr6f_{ABC}\Psi^A\Psi^B\Psi^C\right)
\punkt\cr
}\eqn
$$
The standard superspace calculation (``solving the Bianchi
identities'') 
reveals a sequence of ghost number 0
superfields, constrained to be related by successive fermionic
covariant derivatives: 
$$
\eqalign{
&D_\a A_\b+D_\b A_\a+\{A_\a,A_\b\}+2\g_{\a\b}^iA_i=0\komma\cr
&F_{a\a}=\*_aA_\a-D_\a A_a+[A_a,A_\a]=(\g_a\chi)_\a\komma\cr
&{\cal D}_\a\chi^\b\equiv D_\a\chi^\b+\{A_\a,\chi^\b\}
          =\fr2(\g^{ij})_\a{}^\b F_{ij}\komma\cr
&{\cal D}_\a F_{ab}\equiv D_\a F_{ab}+[A_\a,F_{ab}]=2(\g_{[a}\eta_{b]})_\a\komma\cr
&\ldots\cr
}\Eqn\YMBIEqs
$$
Here, $\chi^\a$ is the physical fermion superfield, and $A_a$ the
gauge connection superfield. 
All fields will of course be on shell. The field denoted $\eta_a^\a$ is
the covariant derivative of $\chi^a$: $\eta_a^\a={\cal D}_a\chi^\a$,
which is $\g$-traceless on shell.

As discussed in the Introduction, we need
to interpret these fields as ghost number 0 parts of pure spinor
superfields. In order to find such fields, we need to reinterpret the
equations (\YMBIEqs) as equations where one power of $\l$ has been
stripped off, not two. 
Contracting with {\it one} $\l$, and again denoting $(\l A)=\Psi$ gives:
$$
\eqalign{
&D_\a\Psi+QA_\a+\{\Psi,A_\a\}+2(\g^i\l)_\a A_i=0\cr
&\*_a\Psi-QA_a+[A_a,\Psi]=(\l\g_a\chi)\cr
&Q\chi^\a+\{\Psi,\chi^\a\}
          =-\fr2(\g^{ij}\l)^\a F_{ij}\komma\cr
&QF_{ab}+[\Psi,F_{ab}]=2(\l\g_{[a}\eta_{b]})\komma\cr
&\qquad\ldots\cr
}\Eqn\PhysFieldsEq
$$
Note that (as soon as one gets to gauge covariant fields) these
identities are non-linear versions of cohomology modulo shift symmetry
described in Section \Shift.

We would like to interpret these as relations with the entire field
$\Psi$, not only the ghost number 0 part. Therefore, we need to replace
the ghost number zero superfields with pure spinor superfields,
constructed as some operators with ghost number $-1$
acting on $\Psi$, such that the
equations (\PhysFieldsEq) hold when $Q\Psi+\Psi^2=0$. 
We thus make an Ansatz
$$
\eqalign{
A_\a&=\hat A_\a\Psi\komma\cr
A_a&=\hat A_a\Psi\komma\cr
\chi^\a&=\hat\chi^\a\Psi\komma\cr
F_{ab}&=\hat F_{ab}\Psi\komma\cr
&\ldots\cr
}\Eqn\FieldsOperators
$$
Note that, since $\Psi$ is fermionic, the operators have opposite
statistics to the corresponding fields.
An almost immediate inspection gives at hand that the equations
(\PhysFieldsEq) follow from $Q\Psi+\Psi^2=0$ given that the operators
satisfy
$$
\eqalign{
[Q,\hat A_\a]&=-D_\a-2(\g^i\l)_\a\hat A_i\komma\cr
\{Q,\hat A_a\}&=\*_a-(\l\g_a\hat\chi)\komma\cr
[Q,\hat\chi^\a]&=-\fr2(\g^{ij}\l)^\a\hat F_{ij}\komma\cr
\{Q,\hat F_{ab}\}&=2(\l\g_{[a}\hat\eta_{b]})\komma\cr
\ldots\cr
}\Eqn\Sequence
$$
It was therefore consistent to make a linear Ansatz.
These expressions can be solved sequentially. Note that the shift term
(in the sense defined in Section \Shift)
occurring in the equation stating the closedness of one operator
defines the next one, again modulo a shift transformation.
The operator equation can be solved explicitly, but, as usual when it
comes to constructing operators with negative ghost number, one needs
to use non-minimal pure spinor variables. The result is
$$
\eqalign{
\hat A_\a&=-(\l\lb)^{-1}\left[\fr8(\g^{ij}\lb)_\a N_{ij}
       +\fr4\lb_\a N\right]\komma\cr
\hat A_a&=-\fr4(\l\lb)^{-1}(\lb\g_aD)
       +\fr{32}(\l\lb)^{-2}(\lb\g_{a}{}^{ij}r)N_{ij}\komma\cr
\hat\chi^\a&=\fr2(\l\lb)^{-1}(\g^i\lb)^\a\*_i
            -\fr{192}(\l\lb)^{-2}(\lb\g^{ijk}r)(\g_{ijk}D)^\a\cr
         &\eqskip-\fr{64}(\l\lb)^{-3}(\g_i\lb)^\a(r\g^{ijk}r)N_{jk}
        \komma\cr
\hat F_{ab}&=\fr8(\l\lb)^{-2}(\lb\g_{ab}{}^ir)\*_i
              +\fr{32}(\l\lb)^{-3}(r\g_{ab}{}^ir)(\lb\g_iD)\cr
          &\eqskip-\fr{256}(\l\lb)^{-4}(\lb\g_{abi}r)(r\g^{ijk}r)N_{jk}
\komma\cr
\ldots&\cr
}\Eqn\ExplicitOperators
$$
(Here, $N^{ab}=(\l\g^{ab}w)$ and $N=(\l w)$, with $w_\a={\*\over\*\l^\a}$.)
These expressions are not strictly unique. Each operator should be
considered as a representative of an equivalence class, modulo exact
terms and shift terms.
Some spinor identities that are useful in this calculation are given
in Appendix B. Appendix C contains information on polynomials in
$\lb$ and $r$ which has also been of help.

Some later calculations can be simplified by noting that the physical
fermion operator can be written
$$
\hat\chi^\a=\fr2(\l\lb)^{-1}(\g^i\lb)^\a\Delta_i\komma\Eqn\ChiInTermsOfDelta
$$
where the modified derivative $\Delta_a$ is defined as
$$
\Delta_a=\*_a+\fr4(\l\lb)^{-1}(r\g_aD)-\fr{32}(\l\lb)^{-2}(r\g_{aij}r)N^{ij}
\punkt\eqn
$$
In addition, the field strength operator is expressible in terms of
the fermion operator, and therefore in terms of $\Delta$, as
$$
\hat F_{ab}=-\fr4(\l\lb)^{-1}(r\g_{ab}\hat\chi)
=\fr8(\l\lb)^{-2}(\lb\g_{ab}{}^ir)\Delta_i\punkt\eqn
$$
The $Q$-closedness modulo shifts of the physical operators is
guaranteed by the BRST transformation of $\Delta_a$:
$$
[Q,\Delta_a]=\fr2(\l\lb)^{-1}(r\g_a\g^i\l)\Delta_i\punkt\eqn
$$


It is straightforward to reverse the whole argument and verify that
when these operators act on a field $\Psi=(\l A)$ satisfying
$Q\Psi+\Psi^2=0$, 
they will give the physical superfields (modulo shifts).
It is important to realise that the on-shell 
meaning of \eg\ $\hat F_{ab}\Psi$ will depend on which the equations
of motion are. Consider for example the difference between the
abelian and non-abelian case in eq. (\FieldsOperators). 
When we, in the following Section, deform the equations
of motion, the on-shell meaning of these fields will obviously change.

\subsection\OperatorIdentities{Operator identities}It is possible to
take the correspondence one step further even 
off-shell, using the
defining properties (\Sequence) of the operators. When we use an
operator like $\hat F_{ab}$ 
in the construction of some
term in the action, it is important to verify that they are
cohomologically equivalent to a derivative together with some other
operator. Otherwise, the counting of deformations may go wrong.
Let us start with the next-to-lowest operator, $\hat A_a$. It should
in some sense be equivalent with a fermionic derivative on $\hat
A_\a$, since this is how the physical fields behave. In order to show
this, we may use the shift symmetries. As mentioned, a shift
transformation of one operator in the sequence leads to a BRST
transformation of the following.
Consider a finite shift of $\hat A_\a$ of the form
$$
\d\hat A_\a=c(\g^i\l)_\a(\hat A\g_i\hat A)\komma\eqn
$$
where $c$ is a dimensionless constant.
In order to calculate the corresponding change in $\hat A_a$ (modulo a
shift), we insert the first relation in eq. (\Sequence) to obtain
$$
[Q,\d\hat A_\a]=c(\g^i\l)_\a\left[
-2(D\g_i\hat A)-2(\g^j\l)_\b \hat A_j(\g_i\hat A)^\b
-2(\g_i\hat A)^\b(\g^j\l)_\b\hat A_j
\right]\Eqn\IntermediateChange
$$
The operator $\hat A_\a$ satisfies 
$$
\eqalign{
(\l\hat A)&=N\komma\cr
(\l\g_{ab}\hat A)&=N_{ab}\punkt\cr
}\Eqn\LambdaAHatRelations
$$
Before these relations can be used, we need to order the operators in
eq. (\IntermediateChange). To this end, we use the relations
$$
\eqalign{
[\hat A_\a,\l^\b]&=(\l\lb)^{-1}\left[\fr8(\g^{ij}\lb)_\a(\g^{ij}\l)^\b
                     -\fr4\lb_\a\l^\b\right]\komma\cr
[\hat A_a,\hat A_\a]&=(\l\lb)^{-1}\lb_\a\hat A_a\komma\cr
}\eqn
$$
which are derived from the explicit expressions (\ExplicitOperators)
(we have a feeling that the second of these relations should have some
more immediate interpretation, but are not aware of one). Using these
commutators to order the terms in eq. (\IntermediateChange) to enable
use of eq. (\LambdaAHatRelations), we obtain, after some calculation, 
a simple expression for
the trivial change in $\hat A_a$:
$$
(\g^i\l)_\a\d\hat A_i=-2c(\g^i\l)_\a\left[
(D\g_i\hat A)+(4N+16)\hat A_i
\right]\Eqn\HatAHatARelation
$$ 
So, anything proportional to the expression inside the square brackets
is a trivial (BRST transformation plus shift) change of $\hat
A_i$. It is straightforward to perform the consistency check that the
right hand side of eq. (\HatAHatARelation) is closed due to
eq. (\Sequence).
Eq. (\HatAHatARelation) agrees with what is stated in the
linearisation of the first equation in (\YMBIEqs).  
This relation has now been lifted in a precise way to the
operator level. Most importantly,
the operator relation is valid off-shell.

This procedure can be continued to the higher-dimensional operators. 
A shift in $\hat A_a$ with 
$$
\d\hat A_a=c(\l\g_a\g^i\hat A)\hat A_i\eqn
$$
leads to 
$$
\d\hat\chi^\a=c\left[
\*_i(\g^i\hat A)^\a-(\g^iD)^\a\hat A_i-(2N+10)\hat\chi^\a
\right]\komma\eqn
$$
where we in the process needed to derive the relation
$(\l\g_a\g^i\hat A)(\g_i\l)_\a=(\g_a\l)_\a(2N+10)$. Again, the
trivial expression in the square brackets matches the linearised
equation for the field $\chi^\a$.

Similar identities, relating the composition of one operator and a
spinorial derivative to the following operator, can be derived
analogously. 

Shifts involving $\hat A_a$ instead of $\hat A_\a$
directly give compositions with bosonic derivatives. Let us take one
example, which also clarifies a technical issue with the shift
symmetry we have yet refrained from mentioning.
Consider a shift in $\hat\chi^\a$ of the form
$$
\d\hat\chi^\a=c(\g^{ij}\l)^\a\hat A_i\hat A_j\punkt\eqn
$$
In order to derive the ensuing trivial change in $\hat F_{ab}$, we
need two interesting relations, namely,
$$
\eqalign{
[\hat A_a,(\l\g_b\hat\chi)]&=-\hat F_{ab}\komma\cr
\hat A^i(\l\g_i\hat\chi)&=b-\*^i\hat A_i\komma\cr
}\eqn
$$
where $b$ is the ``$b$-ghost'', the gauge fixing operator 
[\BerkovitsNonMinimal] associated
with Siegel gauge,
$$
\eqalign{
b&=-\fr2(\l\lb)^{-1}(\lb\g^iD)\*_i
+\fr{16}(\l\lb)^{-2}(\lb\g^{ijk}r)\left(N_{ij}\*_k+\fr{24}(D\g_{ijk}D)\right)\cr
&\qquad-\fr{64}(\l\lb)^{-3}(r\g^{ijk}r)(\lb\g_iD)N_{jk}
-\fr{1024}(\l\lb)^{-4}(\lb\g^{ijm}r)(r\g^{kl}{}_mr)N_{ij}N_{jk}
\punkt\cr}\eqn
$$ 
This gives
$$
[Q,\d\hat\chi^\a]=c(\g^{ij}\l)^\a(2\*_i\hat A_j-\hat F_{ij})
-2c\l^\a(b-\*^i\hat A_i)\punkt\Eqn\BGhostShift
$$
The trivial change in $\hat F_{ab}$ is what one expects, but there is
also a shift generated corresponding to a scalar. We see that it
corresponds to a gauge degree of freedom. In fact, the gauge invariant
entity $(\l\g_a\hat\chi)$, appearing on the right hand side in
eq. (\Sequence), allows not only for a shift
$\d\hat\chi^\a=(\g^{ij}\l)^\a\hat\xi_{ij}$, but also
$\d\hat\chi^\a=\l^\a\hat\xi$ (in fact, there is no shift invariant
expression invariant under the first but not under the second
transformation). While the first of these shifts
corresponds to the presence of the field strength at level $\theta$ in
the superfield $\chi^\a$, the second has nothing to do with physical
degrees of freedom, but corresponds to longitudinal modes. 
The second term in eq. (\BGhostShift), and therefore the gauge type of
shift in $\hat\chi^\a$, is necessary; the first term on the right hand
side is not closed.
In the
process, we have shown that Siegel gauge implies Lorenz gauge.  

We have not checked higher operators explicitly.
Our impression is that there are no single-derivative operators that
are cohomologically equivalent to $\*\hat\chi$ or $\*\hat F$. This
would be as well, since the operators we have formulated (together
with bosonic derivatives) suffice to
write operators corresponding to all gauge-covariant physical fields
in the theory. 

One further comment on the form of the operators. Since any cohomology
can be represented as a function of the minimal variables only
(independent of $\lb$ and $\theta$), one may be tempted to think that
the $r$-independent terms of the operators are the relevant ones, and
that \eg\ $\hat F$ is trivial. This is not the case. The operators map
cohomology to cohomology, 
but they do not respect a gauge choice where
there is no dependence on the non-minimal sector. Indeed, if one
examines the action of $\hat A_a$ on the zero mode
$\Psi=-(\l\g^i\theta)A_i$, one
will find that the two terms contribute equally to the cohomology. In
$\hat\chi^\a$, only $r$-dependent terms contribute to the zero mode,
etc. This is important to understand, so that one does not draw a
conclusion implying that deformation terms containing $\hat F$
(and, consequently, having no $r$-independent terms) would be
trivial. The non-triviality of $\hat F_{ab}$ is of course demonstrated
by the above calculations.



The identities
$$
\eqalign{
[\hat\chi^\a,\hat\chi^\b]&=0\komma\cr
[\hat\chi^\a,\hat F_{ab}]&=0\komma\cr
\{\hat F_{ab},\hat F_{cd}\}&=0\cr
}\Eqn\OperatorsCommuteI
$$
are straightforward to prove. 
Also the higher operators (anti-)commute. 
We also have
$$
\eqalign{
[\hat\chi^\a,(\l\g_a\hat\chi)]&=0\komma\cr
[(\l\g_a\hat\chi),(\l\g_b\hat\chi)]&=0\komma\cr
[(\l\g_a\hat\chi),\hat F_{bc}]&=0\punkt\cr
}\Eqn\OperatorsCommuteII
$$
In addition, $\hat\chi^\a$ is pure, $(\hat\chi\g^a\hat\chi)=0$.
We also have an identity
$$
(\l\g^i\hat\chi)\hat F_{ai}=0\punkt\eqn
$$
These identities are very helpful, and simplify calculations
needed to check the master equation. 

The construction in this Section has been performed for $D=10$
super-Yang--Mills theory. It is of course straightforward to adapt it
to $D=4$, $N=4$ SYM, where also some scalar fields become available as
physical operators. An analogous construction should exist for $D=11$
supergravity [\PureSG]. There, the ``vielbein field'' $\Phi^a$,
possessing a certain shift symmetry, was
already given as an operator of ghost number $-2$ acting on the basic
field $\Psi$. Pure spinor superfields of ghost number 0 could be
derived by further action of ghost number $-1$ operators on
$\Phi^a$. These fields may then be used in higher-derivative terms
along the same lines as in the following Section.


\section\Deformations{Born--Infeld and other deformations}Any term in
the action should be expressed as an integral over all variables,
where the integrand has ghost number 3 and dimension 0. Terms with
higher derivatives of course carry some factors of $\a'$ to match
dimension, but ghost number has to be respected.

A higher-derivative deformation should not change how gauge symmetry
transforms physical component fields. But in a BV framework, gauge
symmetries and equations of motion are inextricably integrated --- the
gauge symmetry of the antifields is the equations of motion for the
fields and vice versa. 
Our basic field $\Psi$ is self-conjugate with respect to the
antibracket, so fields and antifields can not be separated.
In order to ensure that a deformation starts
with the equations of motion for the physical fields, \ie, at order
$\l^2$, one may look for a representative of the deformation
containing exactly two explicit powers of $\l$. 
Any term should respect shift symmetry, and it will be ensured
by the explicit powers of $\l$. On the other hand, an expression which
is a product of shift-invariant expressions runs the risk of being
trivial in view of eq. (\Sequence). As we will see, this gives severe
restrictions --- non-trivial terms can arise when shift symmetry is
achieved by the ``sharing'' of $\l$'s between more than one field.
Another principle, which we believe is true, although we do not have a
strict proof, is that it is sufficient to consider expressions where
not more than one operator acts on each field. 

We will examine the deformation starting at $\a'{}^2$, \ie, ``$F^4$
terms'' and associated higher derivative terms, 
in the abelian and non-abelian settings.

\subsection\AbelianDeformation{The abelian case}Let us first 
consider abelian (Maxwell) theory. The $\a'{}^2$ term
is known, and given, at the level of equations of motion, by
eq. (\OldDeformation). We now follow our program of replacing
superfields by pure spinor superfields, obtained by letting the
physical operators of Section \Shift\ act on $\Psi$.

The lowest
order deformation of the Maxwell action then is
$$
S=S_2+S_4=\int[dZ]\left(\fr2\Psi Q\Psi
+\Fr k4\Psi(\l\g^i\hat\chi)\Psi(\l\g^j\hat\chi)\Psi\hat
F_{ij}\Psi\right)
\Eqn\AbelianAction
$$
where $k$ is a constant proportional to $\a'{}^2$.
It leads to the equations of motion
$$
0=Q\Psi+k(\l\g^i\hat\chi)\Psi(\l\g^j\hat\chi)\Psi\hat F_{ij}\Psi\punkt\eqn
$$
Alternatively, the 4-point coupling can be written
$$
S_4=\Fr
k4\int[dZ]g_{\a\b\g}\Psi\hat\chi^\a\Psi\hat\chi^\b\Psi\hat\chi^\g\Psi
\komma\eqn
$$
where $g_{\a\b\g}=\fr4(\g^i\l)_{[\a}(\g^j\l)_\b(\g_{ij}r)_{\g]}$.
Here, the commutativity of
eqs. (\OperatorsCommuteI) and (\OperatorsCommuteII) 
has been used. 
An even simpler version, using eq. (\ChiInTermsOfDelta), is
$$
S_4=\Fr k{32}\int[dZ](\l\lb)^{-2}(\lb\g^{ijk}r)
\Psi\Delta_i\Psi\Delta_j\Psi\Delta_k\Psi\komma\Eqn\SimplestSFour
$$
We
see that this gives the linear deformation (\OldDeformation) described in refs. 
[\CederwallNilssonTsimpisI,\CederwallNilssonTsimpisII].

We would like to comment on the closedness and non-triviality of the
term $S_4$. It is easily seen that there are no possible trivial terms 
$(S_2,R)$ resulting in $\l^2\Psi\chi^2 F$ terms in the abelian case.
We note that the action (\AbelianAction) has been written in a form
where the $\hat\chi$'s enter in their shift-invariant form $(\l\g^a\hat\chi)$,
which is closed, but not exact. The $\hat F$ factor, on the other hand, is
not paired with a $\l$. The shift-invariant combination with $\hat F$,
$(\g^{ij}\l)^\a\hat F_{ij}$, is BRST-exact, due to eq. (\Sequence), 
so having one $\l$ for
each operator would lead to a trivial expression. Associating the $\l$'s
with the $\hat\chi$'s to form shift-invariant expressions 
is however a matter of taste. One could equally well
write $S_4$ as proportional to 
$$
\Psi(\l\g^i\hat\chi)\Psi(\hat\chi\Psi\g_i[Q,\hat\chi]\Psi)
\komma\eqn
$$ 
by instead associating one of the $\l$'s with $\hat F$, but in the
process leaving one of the $\chi$'s ``naked''.
The key to $S_4$ being closed but not exact is the fact
that the $\l$'s are ``shared'' between fields to guarantee shift
symmetry. The invariance of $S_4$ on the form (\SimplestSFour) is less
obvious, and of course relies on the transformation of the ``prefactor''.


\Yboxdim4pt

It turns out that not only does the action (\AbelianAction) 
define a consistent infinitesimal deformation, \ie, $(S_2,S_4)=0$, but
it also 
fulfills the full master
equation. The identity $(S_4,S_4)=0$ can be seen by writing it
as
$$
\eqalign{
(S_4,S_4)&=k^2\int[dZ](\l\g^a\hat\chi)\Psi(\l\g^b\hat\chi)\Psi\hat F_{ab}\Psi
  (\l\g^c\hat\chi)\Psi(\l\g^d\hat\chi)\Psi\hat F_{cd}\Psi\cr
&=-\Fr{k^2}{64}\int[dZ](\l\lb)^{-4}(\lb\g^{abc}r)(\lb\g^{ijk}r)
\Delta_a\Psi\Delta_b\Psi\Delta_c\Psi\Delta_i\Psi\Delta_j\Psi\Delta_k\Psi
\punkt}\Eqn\SFourSFour
$$
The only non-vanishing modules in $\lb^2r^2$ are
$(01011)\oplus(00110)$ (see Appendix C). The first of these is the one
with six vector indices, a traceless tensor with symmetry 
\lower6pt\hbox{\yng(2,2,1,1)}.
The completely antisymmetric tensor, demanded in eq. (\SFourSFour),
does not occur, so that expression vanishes identically.

The conclusion is that the action (\AbelianAction), with only a
4-point coupling in addition to the kinetic term, actually encodes the
full abelian Born--Infeld theory. This may be seen as somewhat
surprising, but is very much in line with the simplifications
generically occurring in interacting pure spinor field theories
[\CederwallThreeConf,\PureSG]. It will be interesting, and
maybe also instructive as a toy model for supergravity, to try to find
the most efficient way of extracting the
Born--Infeld dynamics for the physical component fields from the
polynomial action (\AbelianAction). In both cases, the non-linearities
generate square roots of determinants.



\subsection\NonAbelianDeformation{The non-abelian case}How does the 
master equation work in the non-abelian case? 
We now depart from the action (\CSAction).
A 4-point
coupling can be written as
$$
S_4=\fr4t_{ABCD}\int[dZ]\Psi^A(\l\g^i\hat\chi)\Psi^B(\l\g^j\hat\chi)\Psi^C
                \hat F_{ij}\Psi^D\komma\eqn
$$
where $t_{ABCD}=t_{(ABCD)}$ is a symmetric invariant tensor in adjoint
indices. 
It is easily shown, by expressing the operators in terms of the
$\Delta$ operator, that other symmetries in $t_{ABCD}$ do not occur. 

The experience from Maxwell theory indicates that
$S=S_{\hbox{\sixrm CS}}+S_4$ satisfies the master
equation, apart from resulting terms which are not completely symmetric in the
adjoint indices. Remaining terms, which contain antisymmetrisations in
adjoint indices, must be compensated by $(S_2,S_6)$.
$(S_2,S_4)$ is calculated as in the
abelian case. $(S_3,S_4)=0$ encodes the invariance of
$t_{ABCD}$, \ie, the identity
$f_{A(B}{}^Ft_{CDE)F}=0$. 
The calculation for $(S_4,S_4)$
works similarly as is the abelian case, but unlike in the abelian case,
we can not assume that the tensor
$t_{ABC}{}^Gt_{DEFG}$ is completely symmetric. 

The result now is
$$
\eqalign{
(S_4,S_4)&=-\fr{64}\int[dZ]t_{ABC}{}^Lt_{IJKL}
(\l\lb)^{-4}(\lb\g^{abc}r)(\lb\g^{ijk}r)\cr
&\qquad\qquad\times
\Delta_a\Psi^A\Delta_b\Psi^B\Delta_c\Psi^C
\Delta_i\Psi^I\Delta_j\Psi^J\Delta_k\Psi^K\komma\cr}\Eqn\SFSFNonAb
$$
and we notice that this projects on the
symmetry \lower2pt\hbox{\yng(4,2)} in the tensor
$t_{ABC}{}^Gt_{DEFG}$. The right hand side of eq. (\SFSFNonAb) is not
a total derivative.
Six-point (and presumably higher) terms are
needed. We have not calculated such additional terms, and have so far not
been able to deduce their structure using the properties of the
operators in Section \Operators.
We note that the entire contribution with symmetrised traces is
encoded in $S_4$.


\section\Conclusions{Conclusions and outlook}We have presented a
method for lifting linear deformations constructed earlier in the
superspace/pure spinor formulation to a consistent BV framework. The
construction has a simple intuitive meaning in terms of operators
corresponding to physical
fields. These operators turned out to have remarkably simple
properties. Most notably, the by now quite common phenomenon that pure
spinor superspace BV actions reduce the degree of interactions is at
work, to the degree that the abelian Born--Infeld action to all orders
is represented by a kinetic term and a 4-point interaction. The
non-abelian deformation also simplifies, in that the 4-point term
encodes the full contribution from symmetrised traces.

Many question arise, which we have not yet addressed:

How are the equations of
motion for the component field extracted?
Even though all (linear) cohomology has representatives independent of
the non-minimal variables $\lb$ and $r$, this will not be true for the
interacting theory. The minimal form of the Born--Infeld superspace
equations of motion obtained in ref. [\BerkovitsPershinBI] might serve
as a guideline. 
We hope that a systematic investigation of how Born--Infeld component
equations of motion are reproduced can shed some light also on the
nature of the polynomial action for $D=11$ supergravity [\PureSG].


It is conceivable that our framework is efficient for finding more
general supersymmetric invariants. It would be interesting to
investigate this issue, both for $D=10$ SYM and for its reduction to
$D=4$, where more possibilities may arise, and compare with 
refs. [\MovshevDef,\MovshevSchwarzDef].


A similar construction should be performed in $D=11$ and $D=4$, $N=8$
supergravity. We would also like to understand how to implement
U-duality in our formalism, since it seems to have an important r\^ole
to play in the investigation of supergravity counterterms (for recent
considerations of invariants in supergravity, see \eg\
refs. [\BHLSW,\BeisertEtAlSG]).



\appendix{Conventions and notation}All calculations are performed in
flat space or flat superspace. Lorentz indices are denoted
$a,b,\ldots$ or $i,j,\ldots$, while chiral spinor indices are
$\a,\b,\ldots$. Lie algebra (adjoint) indices are $A,B,\ldots$

Flat superspace covariant derivatives in inertial basis are denoted
$D_\a$ (fermionic) and $\*_a$ (bosonic).

Contractions of spinor indices are denoted $(...)$, \eg\ 
$(\l\g^a\chi)\equiv\l^\a\g^a_{\a\b}\chi^\b$.

Commutators and anticommutators of operators and fields are denoted
with the same symbols, $[\cdot,\cdot]$ and $\{\cdot,\cdot\}$
respectively. The distinction should be clear from the context. 
When this denotes a commutator of fields (two
bosonic or one bosonic and one fermionic) or an anticommutator of two
fermionic fields, the notation is shorthand for (in the first case) 
$[U,V]^A\equiv f^A{}_{BC}U^BV^C$, not only for matrix
algebras. Similarly the square of a fermionic Lie algebra valued field
means $(\Psi^2)^A\equiv\fr2f^A{}_{BC}\Psi^B\Psi^C$.

Batalin--Vilkovisky antibrackets are denoted $(\cdot,\cdot)$.

\appendix{Some useful identities}The antisymmetric product of three
spinors is a $\g$-traceless 2-form spinor of the opposite chirality,
$\wedge^3(00001)=(01010)$. This is manifested in the identity
$$
\theta^\a(\theta\g_{abc}\theta)=\fr2(\g_{[a}\g^i\theta)^\a(\theta\g_{bc]i}\theta)
\komma\eqn
$$
where the normalisation is determined by taking the $\g$-trace.

Some useful relations involving the pure spinor $\l$ and the invariant
combinations $N^{ab}\equiv(\l\g^{ab}w)$ and $N\equiv(\l w)$ are:
$$
\eqalign{
&(\g^j\l)_\a N_{ij}=(\g_i\l)_\a N\komma\cr
&(\g^{ij}\l)^\a N_{ij}=10\l^\a N\komma\cr
&(\g^{ij}\g_{abc}\l)_\a N_{ij}=-2(\g_{abc}\l)_\a N-24(\g_{[a}\l)_\a
N_{bc]}\punkt\cr
}\eqn
$$

Some of the algebraic calculations involving spinors have been
facilitated by the use of the Lie algebra program LiE [\LiE] and
the Mathematica package GAMMA [\GAMMA].

\appendix{Partition functions for the non-minimal variables}In many of
the calculations, it is practical to know the Lorentz modules
appearing in some monomial $\lb^nr^m$. The LiE code given at the end
of this appendix defines a function that does precisely this.
The result may be summarised in a table:

\vskip3\parskip
\thicksize=\thinsize
\ruledtable
|$\lb^0$|$\lb^1$|$\lb^{n\geq2}$\cr
$r^0$|$\ss(00000)$|$\ss(00010)$|$\ss(000n0)$\cr
$r^1$|$\ss(00010)$|$\ss(00020)\oplus(00100)$|$\ss(000,n+1,0)\oplus(001,n-1,0)$\cr
$r^2$|$\ss(00100)$|$\ss(00110)\oplus(01001)$|$\ss(001n0)\oplus(010,n-1,1)$\cr
$r^3$|$\ss(01001)$|$\ss(01011)\oplus(02000)\oplus(10002)$
                |$\ss(010n1)\oplus(020,n-1,0)\oplus(100,n-1,2)$ \cr
$r^4$|$\ss(02000)\oplus(10002)$|$\ss(00003)\oplus(02010)\oplus(10012)\oplus(11001)$
          |$\ss(000,n-1,3)\oplus(020n0)\oplus(100n2)\oplus(110,n-1,1)$\cr
$r^5$|$\ss(00003)\oplus(11001)$|$\ss(00013)\oplus(01002)\oplus(11011)\oplus(20100)$
     |$\ss(000n3)\oplus(010,n-1,2)\oplus(110n1)\oplus(201,n-1,0)$\cr
$r^6$|$\ss(01002)\oplus(20100)$|$\ss(01012)\oplus(10101)\oplus(20110)\oplus(30010)$
      |$\ss(010n2)\oplus(101,n-1,1)\oplus(201n0)\oplus(300n0)$\cr
$r^7$|$\ss(10101)\oplus(30010)$
        |$\ss(00200)\oplus(10111)\oplus(20011)\oplus(30020)\oplus(40000)$
        |$\ss(002,n-1,0)\oplus(101n1)\oplus(200n1)\oplus(300,n+1,0)$\cr
$r^8$|$\ss(00200)\oplus(20011)\oplus(40000)$
       |$\ss(00210)\oplus(10110)\oplus(20021)\oplus(30001)$
       |$\ss(002n0)\oplus(101n0)\oplus(200,n+1,1)$\cr
$r^9$|$\ss(10110)\oplus(30001)$
        |$\ss(01020)\oplus(10120)\oplus(20100)$
        |$\ss(010,n+1,0)\oplus(101,n+1,0)$\cr
$r^{10}$|$\ss(01020)\oplus(20100)$|$\ss(00030)\oplus(01030)\oplus(11010)$
         |$\ss(000,n+2,0)\oplus(010,n+2,0)$\cr
$r^{11}$|$\ss(00030)\oplus(11010)$|$\ss(00040)\oplus(02000)\oplus(10020)$
        |$\ss(000,n+3,0)$\cr
$r^{12}$|$\ss(02000)\oplus(10020)$|$\ss(01010)$|$\ss\emptyset$\cr
$r^{13}$|$\ss(01010)$|$\ss(00100)$|$\ss\emptyset$\cr
$r^{14}$|$\ss(00100)$|$\ss(00001)$|$\ss\emptyset$\cr
$r^{15}$|$\ss(00001)$|$\ss(00000)$|$\ss\emptyset$\cr
$r^{16}$|$\ss(00000)$|$\ss\emptyset$|$\ss\,\,\,\emptyset$
\endruledtable
\vskip2\parskip

We define partition functions $P_m(t)$
counting the number of states at a given power $r^m$ and arbitrary
power of $\lb$. $P_0$ is the usual pure spinor partition function. The
complete set of functions is
$$
\eqalign{
P_0(t)&=(1-t)^{-11}(1+5t+5t^2+t^3)\komma\cr
P_1(t)&=(1-t)^{-11}(16 + 70 t + 46 t^2)\komma\cr
P_2(t)&=(1-t)^{-11}(120 + 440 t + 110 t^2 - 10 t^3)\komma\cr
P_3(t)&=(1-t)^{-11}(560 + 1600 t - 416 t^2 + 446 t^3 - 330 t^4 + 165
t^5 \cr 
&\eqskip- 55 t^6 +  
 11 t^7 - t^8)\komma\cr
P_4(t)&=(1-t)^{-11}(1820 + 3500 t - 3460 t^2 + 4620 t^3 - 4620 t^4 + 3300 t^5 \cr 
&\eqskip- 
 1650 t^6+ 550 t^7 - 110 t^8 + 10 t^9)\komma\cr
P_5(t)&=(1-t)^{-11}(4368 + 3640 t - 9800 t^2 + 20608 t^3 - 29040 t^4 + 28578 t^5 \cr 
&\eqskip- 
 19910 t^6+ 9680 t^7 - 3136 t^8 + 610 t^9 - 54 t^{10})\komma\cr
P_6(t)&=(1-t)^{-11}(8008 - 66440 t + 269434 t^2 - 656238 t^3 + 1016400 t^4 - 965184 t^5 \cr 
&\eqskip+ 
 408870 t^6 + 233662 t^7 - 497816 t^8 + 373320 t^9 - 156658 t^{10} \cr 
&\eqskip+ 
 36470 t^{11} - 3696 t^{12})\komma\cr
P_7(t)&=(1-t)^{-11}(11440 - 22880 t + 20800 t^2 + 51480 t^3 - 231000 t^4 + 456060 t^5 \cr 
&\eqskip- 
 581460 t^6 + 518870 t^7 - 329890 t^8 + 147350 t^9 - 44130 t^{10} \cr 
&\eqskip+ 
 7980 t^{11} - 660 t^{12})\komma\cr
P_8(t)&=(1-t)^{-11}(12870 - 50050 t + 109604 t^2 - 98044 t^3 - 142230 t^4 + 629970 t^5 \cr 
&\eqskip- 
 1093180 t^6 + 1187956 t^7 - 877211 t^8 + 443225 t^9 - 147615 t^{10} \cr 
&\eqskip+ 
 29325 t^{11} - 2640 t^{12})\komma\cr
P_9(t)&=(1-t)^{-11}(11440 - 69718 t + 225698 t^2 - 428120 t^3 + 429000 t^4 + 6270 t^5 \cr 
&\eqskip- 
 705034 t^6 + 1142504 t^7 - 1025320 t^8 + 586190 t^9 - 213290 t^{10} \cr 
&\eqskip+ 
 45352 t^{11} - 4312 t^{12})\komma\cr
P_{10}(t)&=(1-t)^{-11}(8008 - 66440 t + 269434 t^2 - 656238 t^3 + 1016400 t^4 - 965184 t^5 \cr 
&\eqskip+
 408870 t^6 + 233662 t^7 - 497816 t^8 + 373320 t^9 - 156658 t^{10} \cr 
&\eqskip+ 
 36470 t^{11} - 3696 t^{12})\komma\cr
P_{11}(t)&=(1-t)^{-11}(4368 - 43456 t + 199232 t^2 - 544390 t^3 + 970530 t^4 - 1160049 t^5 \cr 
&\eqskip+
 905707 t^6 - 398761 t^7 + 15995 t^8 + 95650 t^9 - 59318 t^{10} \cr 
&\eqskip+ 
 16324 t^{11} - 1820 t^{12})\komma\cr
P_{12}(t)&={1820+560t}\komma\cr
P_{13}(t)&={560+120t}\komma\cr
P_{14}(t)&={120+16t}\komma\cr
P_{15}(t)&={16+t}\komma\cr
P_{16}(t)&=1\komma\cr
}\eqn
$$

The BRST operator acts
``south-west'' in the table. 
It is easily checked explicitly that these modules pair up in a way
consistent with $1$ being the only cohomology. This is of course a
consequence of the 
fact that the purity constraint on $r$ is the BRST variation of the
one on $\lb$.

Consider for example the commutator $[\hat\chi^\a,\hat\chi^\b]$. One
possible term comes from the anticommutator of the fermionic covariant
derivatives. It contains $(\l\lb)^{-4}\lb^2r^2\*$ and transforms as
$(00100)$. Expanding $P_2(t)$ shows that the dimension of the modules
occurring at $\lb^2r^2$ is 12870. A more detailed calculation,
performed by hand, or with the LiE code below, shows that these
modules are $(00120)\oplus(01011)$. None of them contributes to
$(00100)$ when multiplied by a vector $(10000)$, so this term vanishes
for purely representation theoretical reasons. Such arguments in fact
apply to any term in the equations (\OperatorsCommuteI) and
(\OperatorsCommuteII), including
those coming from $N_{ab}$ acting on the $(\l\lb)^{-p}$ prefactors.

\line{\hrulefill}

{\tt
\noindent 
\null \#\#\#\#\#\ some definitions \#\#\#\#\#\       \hfill\break
\null setdefault D5                                  \hfill\break
\null lb=1X[0,0,0,1,0]                               \hfill\break
\null s=1X[0,0,0,1,0]                                \hfill\break
\null v=1X[1,0,0,0,0]                                \hfill\break
\null r(int n)=1X[0,0,0,n,0]                         \hfill\break
\null as(int n)=alt\_tensor(n,s)                     \hfill\break
\null \#\#\#\#\#\ the positive part of a polynomial \#\#\#\#\#\ \hfill\break
\null pos\_pol(pol p) =                              \hfill\break
\null $\{$                                           \hfill\break
\null loc q=p;                                       \hfill\break
\null for i=1 to length(p) do                        \hfill\break
\null\hskip1cm if coef(p,i)<0 then q=q-p[i];         \hfill\break
\null\hskip1cm fi;                                   \hfill\break
\null od;                                            \hfill\break
\null q                                              \hfill\break
\null $\}$                                           \hfill\break
\null \#\#\#\#\#\ modules at r\^{}m lambdabar\^{}n \#\#\#\#\#\  \hfill\break
\null rr(int m,n)=                                   \hfill\break
\null $\{$                                           \hfill\break
\null if m==0 then                                   \hfill\break
\null\hskip1cm 	r(n);                           \hfill\break
\null else                                           \hfill\break
\null\hskip1cm if n==0 then                          \hfill\break
\null\hskip2cm 	as(m);                          \hfill\break
\null\hskip1cm else                                  \hfill\break
\null\hskip2cm pos\_pol(tensor(as(m),r(n))-tensor(v,tensor(as(m-1),r(n-1)))); \hfill\break
\null\hskip1cm fi;                                   \hfill\break
\null fi;                                            \hfill\break
\null $\}$                                           \hfill\break
}
\line{\hrulefill}

\refout

\end